\documentclass{article}[12pt]
\usepackage{amsmath}
\usepackage{amssymb}
\usepackage{amsthm}
\usepackage{amsfonts}

\def\abstract#1{\vskip 7mm
        \begin{center}{\large Abstract}\par \smallskip
                \begin{minipage}[c]{12cm}
                        \small #1
                \end{minipage}
        \end{center}
}

\def\title#1{\begin{center}{\Large\bf #1}\end{center}}
\def\author#1{\vskip 5mm \begin{center}{#1}\end{center}}
\def\address#1{\begin{center}{\it #1}\end{center}}

\newcommand{\ssmatrix}[4]%
{\begin{pmatrix} #1 & #2 \\ #3 & #4 \end{pmatrix}}
\makeatletter
\@ifundefined{lesssim}{}{}
\@ifundefined{gtrsim}{}{}
\def\vereq#1#2{\lower3pt\vbox{\baselineskip1.5pt \lineskip1.5pt
\ialign{$\m@th#1\hfill##\hfil$\crcr#2\crcr\sim\crcr}}}
\makeatother

\newtheorem{theorem}{Theorem}
\newtheorem{definition}{Definition}
\newtheorem{corollary}{Corollary}

\newtheorem{lemma}{Lemma}
\newtheorem{proposition}{Proposition}
\newtheorem{axiom}{Connes' axioms}

\newcommand{\dirac}{\not\!\partial}

\newcommand{\dd}{{\cal D}}
\newcommand{\Hs}{{\cal H}}

\newcommand{\G}{{\cal G}}
\newcommand{\tm}{{\rm T}M}
\newcommand{\alg}{{\cal A}}
\newcommand{\bop}{{\cal L(H)}}
\newcommand{\cop}{{\cal K(H)}}
\newcommand{\lop}{{\cal L(H)}}
\newcommand{\U}{{\mathbb I}}
\newcommand{\tr}{{\rm Tr}}
\newcommand{\Dtr}{{\rm Tr^+}}
\newcommand{\scom}[1]{[#1]_S}
\newcommand{\str}{{\rm Str}}
\newcommand{\RR}{{\mathbb R}}
\newcommand{\CC}{{\mathbb C}}
\newcommand{\NN}{{\mathbb N}}
\newcommand{\ZZ}{{\mathbb Z}}
\newcommand{\Wres}{{\rm Wres}}
\newcommand{\wres}{{\rm wres}}
\newcommand{\Dc}{{\cal L}^{1+}}
\DeclareMathOperator*{\res}{Res}

\newcommand{\Cinf}{\mbox{$C^\infty$}}

\begin{document}

\title{%
Operator geometry and algebraic gravity}
\author{%
  Masaru Siino\footnote{E-mail:msiino@th.phys.titech.ac.jp} 
}
\address{%
  Department of Physics, Tokyo Institute of Technology, \\
  Oh-Okayama 1-12-1, Megro-ku, Tokyo 152-8550, Japan
}

\abstract{
An algebraic formulation of general relativity is proposed.
The formulation is applicable to quantum gravity and noncommutative space.
To investigate quantum gravity we develop the canonical formalism of operator geometry, after reconstructing an algebraic canonical formulation on analytical dynamics. The remarkable fact is that the constraint equation and evolution equation of the gravitational
system are algebraically unified.
From the discussion of regularization
we find the quantum correction of the semi-classical gravity is same as that already known in quantum field theory.
}

\section{Introduction}

There is no  doubt that the essence of quantum theory is the noncommutativity of operator algebra for physical variables.
Usually, this noncommutativity reflects the ordering of quantum observations.
In quantum gravity, how can we understand the noncommutativity?
A standard answer would be the framework of field theory where the gravitational field is decomposed into a fluctuation of potential
and background geometry. As long as the gravity is sufficiently weak, that will promise to treat spacetime geometry.
 However, strong gravity drastically changes
background geometry, since general relativity is remarkably non-linear theory.
Indeed, the topological instanton of quantum gravity implies the microscopic nontriviality of topology for spacetime.
In this sense one should care the topological structure as a physical variable, for example, the Wilson loop and moduli parameter.
If we want to completely enumerate the topological variables, one might treat the open neighborhoods of spacetime points,
 whose set and coordinate transformation yields a concept of manifold. 
Roughly speaking, considering all open neighborhoods as physical variable is similar to regarding their own
spacetime points as physical variables by themselves.
This will realize any noncommutativity through quantization of spacetime dynamics.
Of course, though such an idea is an imaginary picture without mathematical background, in the context of noncommutative geometry, algebraic consideration, that is given by algebraic abstraction of geometry, will realize the concept of noncommutativity of spacetime points!
Then we hope the investigation of the noncommutative geometry sheds light on the dark side of quantum gravity.

The most successful framework of the noncommutative space has been established by A. Connes and is well known by his textbook\cite{NCG}.
In the study of the noncommutative geometry, the most important concept is algebraic abstraction of conventional commutative geometry.
While ordinary commutative geometry has been established on a space with any mathematical structure, noncommutative geometry is constructed on the basis of the operator algebra which acts on a Hilbert space.
Therefore we need algebraic abstraction of gravity theory when we consider  noncommutative gravity theory.
We will have corresponding algebraic objects instead of the conventional
objects like curvatures by which we characterize geometry, that is, topological space, differentiable manifold and Riemannian manifold.
The algebraic counterparts  are a set of continuous function, pseudodifferential  operators and a Dirac operator; and are defined in the framework of a commutative operator algebra.
Once this algebraic abstraction has been made, the operator algebra naturally extends from commutative to noncommutative one.

One might worry that by the noncommutative extension, space may lose its continuity or differentiability. The most worrisome thing might be missing of reasonable measure which is indispensable to develop physical theory there.
Nevertheless, the framework of the spectral geometry established by Connes\cite{88}\cite{469} provides the substantial structure that is appropriate to define a
trace of operators.

Although people believe that our world is commutative and relativists study the dynamics of commutative spacetime, 
 there are two motivations to consider noncommutative space.
One is interest about the characteristic of real noncommutative space.
For example, the noncommutative torus or the space of irreducible Penrose tiling is a well-known model of noncommutative space. The noncommutative torus of irrotational ring\cite{390} is known as not Lebesgue measurable space. 
The set of the Penrose tiling
is  a quotient set of the Cantor set\cite{NCG}.
The other is physical interest concerning quantum gravity.
In the study of quantum theory of gravitation, the string theory is paid the most attention and it seems there has been much progress.
In exchange of that, it forces us uncomfortable changes of our view on the world, for example extra dimensional spacetime, infinite number of vacuum states, and so on.
On the other hand, the outlook of the world in the loop gravity theory might be easier to accept than that of the string theory  but
the quantum state in the loop gravity is hard to physically interpret.
Recently many authors study the quantum space, in the standpoint that at Planckian scale  spacetime should be noncommutative.
One of consistent way, to incorporate the noncommutative quantum space is to formulate gravity in an algebraic manner. 
In the present work, we apply the algebraic method developed by Connes and other people\cite{NCG}\cite{ENG}, to general relativity.

In those viewpoints, it is promising to develop quantum gravity in the context of noncommutative geometry.
The main purpose of the present article is to reformulate general relativity in terms of operator algebra.
One will get a new equation of gravitation as the algebraic equation of bounded operators on a Hilbert space.
Then we advance toward quantum geometry through the canonical formalism and will find that constraint equation can be formally solved. Moreover from a semi-classical approximation to
 algebraic gravity we reproduce the known quantum correction of field theory.
Though the full development of quantization is postponed until forthcoming works\cite{FCW},
it is possible to discuss the ability of the algebraic formulation in study of quantum gravity.


\subsection{topological notion and Connes' axioms}

In a sense, noncommutative geometry is a kind of movement in the mathematical community to change ordinary multiplication which appears in many aspects of analytical geometry into noncommutative multiplication.
To develop the noncommutative geometry, the algebraic abstraction of topological notion is fundamental, 
that is based on the following  Gel'fand-Na\u{i}mark's\cite{194} two theorems.
\begin{theorem}[Gel'fand-Na\u{i}mark 1]
If $\alg$ is a commutative $C^*$-algebra, the Gel'fand  transformation is an isometric $*$-isomorphism of $\alg$ onto
$C_0(M(\alg))$.
\label{thm:gn1}
\end{theorem}
There $M(\alg)$ is the spectrum space which is a set of all characters $\mu$ (the unitarily equivalence class of
irreducible representation) with Gel'fand topology\footnote{the weak* topology for a subset of the dual space of $\alg$}.
Gel'fand transformation of $a\in \alg$ is the function $\hat{a}:M(\alg)\mapsto {\bf C}$ given by
\begin{align}
\hat{a}(\mu):=\mu(a).
\end{align}

If $\alg$ is unital, $M$ is compact.
For instance, the convolution algebra $\alg=L^1({\RR})$ of Lebesgue-integrable functions of $\RR$ is a nonunital algebra
whose characters are the integrals $f\mapsto \int_\RR e^{-itx}f(x)$ for any $t\in\RR$ then $\hat{f}$ is the Fourier transform
of $f$, and the Gel'fand transformation is the Fourier transformation that takes $L^1(\RR)$ into $C_0(\RR)$.
This theorem asserts that $C_0(M)$ can work as a substitute of topological space $M$, because 
all topological information of $M$ is algebraically encoded in $C_0(M)$.
We can even gain some insight of the noncommutative space which is beyond naive analogy to conventional topology by regarding a noncommutative $C^*$-algebra as a kind of function algebra. This viewpoint allows one to study the topology  of non-Hausdorff space with which we are encountered in probing a continuum where points are unresolved.

The second theorem of Gel'fand-Na\u{i}mark states that any $C^*$-algebra can be embedded as a norm-closed subalgebra of a full algebra of operators for a large enough Hilbert space $\Hs$; 
\begin{theorem}[Gel'fand-Na\u{i}mark 2]
Any $C^*$-algebra has an isometric representation as a closed subalgebra of the algebra $\lop$ of bounded operators on some Hilbert space.
\end{theorem}
Thus all $C^*$-algebras gets fairly concrete representation.
Indeed, $C(M)$ may be embedded in $\lop$, the algebra of bounded operators on the Hilbert space $\Hs$ with a countable orthonormal basis, in many ways: if $M$ is infinite but separable, take $\nu$ to be any finite regular Borel measure on $M$, and identify $\Hs$ with $L^2(M,d\nu)$; then $f\in C(M)\subset L^{\infty}(M,d\nu)$ can be identified with the multiplication operator $h\mapsto fh$ on $L^2(M,d\nu)$.
It is in the spirit of noncommutative geometry to relax closure conditions for a Banach algebra as much as possible when defining our algebras. For instance, if $M$ is a compact differentiable manifold, we would like to use the algebra of smooth functions $\alg=C^\infty(M)$, which is a dense subalgebra of $C(M)$.
Of course, we are able to consider noncommutative algebra instead.
Then corresponding space is no longer locally compact or continuous.
These possibilities will not be discussed in detail in the present article, however.


We can see a good guidance  to an extension of
general relativity to noncommutative spacetime in noncommutative gravity.
At least three gravity models are possible.
They are `gravity \`{a} la Connes-Dixmier-Wodzicki'\cite{CDW}, `spectral gravity'\cite{SG}, and `linear connections'\cite{LC}
Former two are essentially same but with different spirits.
In the spirit of spectral gravity, the eigenvalues of the Dirac operator, which are diffeomorphic invariant functions of the
geometry and therefore true observable in general relativity, have been taken as a set of variables for
an invariant description of the dynamics of the gravitational field.
In the present article, we develop the algebraic formulation of general relativity in the sense of
gravity \`{a} la Connes-Dixmier-Wodzicki.
A similar trial was carried by Hawkins\cite{HK}, but in his formulation spacetime concept is not incorporated since only the spacelike
section is noncommutative there.
As one can see later, the noncommutativity in the direction of evolution is the essence of the present work and
 is closely related to quantum theory of geometry.
Our purpose is to establish an algebraically formulated dynamics of geometry.
The scheme to construct gravity models in noncommutative geometry\cite{cmp182}, and in fact to reconstruct the full geometry
out of the algebra $C^\infty(M)$, is based on the  Dirac operator (a spectral triple) regarded as a dynamical variable and the use of the Dixmier trace and Wodzicki residue\cite{490-491}.
 That is, Connes' axioms of noncommutative gravity is presented as
\begin{axiom}
Suppose we have a smooth compact manifold $M$ without boundary and of dimension $n$. Let $\alg=C^\infty(M)$ and $\dd$ just a
`symbol' for the time being. Let $(\alg_{\pi},\dd_\pi)$ be a unitary representation of the couple $(\alg, \dd)$ as operators on an Hilbert space $\Hs_\pi$ such that $(\alg_\pi, \dd_\pi, \Hs_\pi)$ satisfy
 axioms of a spectral triple (also called an unbounded $K$-cycle)\cite{NCG}\cite{ENG}.
Then
\begin{enumerate}
\item
There is a unique Riemann metric $g=g(\dd)$ on $M$, whose geodesic distance between any two points on $M$ is given by
\begin{align}
d_g(p,q)=\sup\{|f(p)-f(q)|:||[\dd,f]||\leq 1, f\in C^\infty(M)\},\ \  \forall p,q\in M
\label{eqn:dpq}
\end{align}
\item
The metric $g$ depends only on the unitary equivalence class of the representations to metrics form a finite collection
of affine spaces $\alg_\sigma$ parametrized by the spin structures $\sigma$ on $M$.
\item
The action functional given by the Wodzicki residue
\begin{align}
S(\dd)\sim \Wres(\dd^{2-n})
\end{align}
is a positive quadratic form with a unique minimum $\pi_\sigma$ on each $\alg_\pi$.
The minimum $\pi_\sigma$ is the representation of $(\alg,\dd)$ on the Hilbert space of square integrable spinors
$L^2(M,S_\sigma)$. That is, $\alg_\sigma$ acts by multiplicative operators and $\dd_\sigma$ is the Dirac operator of the Levi-Civita connection.
\label{enum3}
\item
At the minimum $\pi_\sigma$, the values of $S(\dd)$ coincides with the Wodzicki residue of $\dd^{n-2}_\sigma$ and is proportional to the Einstein-Hilbert action of general relativity
\begin{align}
S(\dd_\sigma)=\Wres(\dd_\sigma^{2-n})=c_n\int_M Rdv\ \ .
\end{align}
\end{enumerate}
\label{axioms}
\end{axiom}

According to the Connes' axioms, we develop the formulation of gravity by operator algebra (`operator geometry')
 and its canonical formalism.
We will find that the constraint equation of gravitational system is algebraically solvable.
Though we do not plunge into the construction of full quantum theory in the present article, in the semi-classical discussion
of regularization, it is clarified that quantum correction of the gravitational action is given, that is the 
already well-known one in quantum field theory of curved space by the heat kernel expansion.

In  section 2, we briefly introduce some elements of noncommutative geometry as an fundamental concepts and
 recapitulate the fundamental frameworks of spectral calculus developed by Connes and other people.
In particular the Connes' axioms are given in this section. In section 3, we develop the canonical formulation of algebraically formulated theory of gravitation based on the section 2, after giving an algebraic abstraction of analytical dynamics.
 The relation between operator geometry and classical gravity is investigated in  section 4 together with
 quantum correction.
The final section is devoted to conclusion and discussions.

The notation related to the noncommutative geometry is according to the Reference\cite{ENG}.

\section{noncommutative geometry and trace of operator}

In this section we briefly review some of the main concepts of noncommutative geometry in order to make this article self contained and to fix our notation. A hasted reader might skip this section   by accepting  Connes' axioms.
For a more comprehensive presentation of this subject we refer to Ref. \cite{ENG}.

\subsection{Dirac operator as spectral triple}
\label{subsec:dirac}
Let $\alg$ be an associative unital algebra. 
We can construct a bigger algebra $\Omega\alg$ out of it by associating to each element $a\in\alg$ a symbol $\delta a$. 
$\Omega\alg$ is the free algebra generated by the symbols $a$, $\delta a$, $a\in \alg$ modulo the relation
\begin{equation}
\delta(ab)=\delta a\ b+a\delta b\ \ .
\end{equation}

With the definition
\begin{align}
\delta(a_0\delta a_1\cdot\cdot\cdot\delta a_k)&:=\delta a_0\delta a_1\cdot\cdot\cdot\delta a_k \ \ ,\\
\delta(\delta a_1\cdot\cdot\cdot\delta a_k)&:=0\ \ ,
\end{align}
$\Omega\alg$ becomes a $\ZZ$-graded differential algebra with the odd differential $\delta$ and $\delta^2=0$.
$\Omega\alg$ is called the universal differential envelope of $\alg$.

The next element in this formalism is a $K$-cycle $(\Hs,\dd)$ over $\alg$, where $\Hs$ is a Hilbert space such that there is an algebra homomorphism
\begin{equation}
\pi : \alg\mapsto \bop ,
\end{equation}
where $\bop$ denotes the algebra of bounded operators acting on $\Hs$. $\dd$ is an unbounded self-adjoint operator with compact resolvent such that $[\dd,\pi(a)]$ is bounded for all $a\in \alg$.
It is the triple $(\alg,\Hs,\dd)$ which contains all geometric information.

We can use $\dd$ to extend $\pi$ to an algebra homomorphism of $\Omega\alg$ by defining
\begin{equation}
\pi(a_0\delta a_1\cdot\cdot\cdot \delta a_k):=\pi(a_0)[\dd,\pi(a_1)]\cdot\cdot\cdot[\dd,\pi(a_k)]\ .
\end{equation}
However, in general $\pi(\Omega\alg)$ fails to be a differential algebra.
In order to repair this, one has to divide $\Omega\alg$ by the two sided ${\ZZ}$-graded differential ideal
$
{\cal J}=\bigoplus_{k\in {\cal N}}{\cal J}^k\ ,\ \ \  {\cal J}^k:=(\ker\pi)^k+\delta(\ker\pi)^{k-1}
$
and we can define the non-commutative generalization of the de Rham algebra, $\Omega_D\alg$, 
$
\Omega_D\alg:=\bigoplus_{k\in{\cal N}} \pi(\Omega^k\alg)/\pi({\cal J}^k)$ as \cite{KPPW}.
If we take, for example, $\alg=C^{\infty}(M)$, the algebra of smooth functions on a compact Riemannian spin manifold $M$, $\Hs$ as the space of square-integrable spin-sections and the Dirac operator $\dd= \dirac$, then $\Omega_D\alg$ is the usual de Rham algebra\cite{NCG}.

Here a remarkable fact of the Connes' first axiom is that the geodesic distance $d(p,q)$ on such a manifold $M$ for any  $p,q\in M$ is encoded in the Dirac operator $\dd$;
\begin{equation}
d(p,q)=\sup_a\{|a(p)-a(q)|;||[\dd,a]||<1,a\in C^\infty(M)\}.
\label{eqn:dis}
\end{equation}
The algebra may extend to a general $C^*$-algebra $\alg$ not required to be commutative, where the positions $p,q$ are
replaced by states of $C^*$-algebra $\psi,\phi$ which is a positive linear functional of norm one.
The reason is because the positions $p,q$ are related to characters $\mu_p,\mu_q$ of $C^\infty(M)$
 by the Gel'fand transformation, which
are just pure states that are not convex combination of other states.
So we expect that the distance function is defined on the full state space which is the closed convex hull of the set of the pure
states.
\begin{equation}
d(\psi,\phi)=\sup_a\{|\psi(a)-\phi(a)|;||[\dd,a]||<1,a\in \alg\}.
\label{eqn:ndis}
\end{equation}
With a representation $(\pi,\Hs)$, the pure states can be related to a vector $|\xi_\psi>$ of $(\pi,\Hs)$ 
by $\psi(a)=<\xi_\psi|\pi(a)\xi_\psi>$.

No information of arcs is involved on the right hand side of this relation and therefore eq.(\ref{eqn:dis}) or (\ref{eqn:ndis}) can be taken as a definition of geodesic distance which still makes sense in situations where arcs cannot be defined.
 We feel this as a motivation to construct a gravitational action which only depends on the choice of the Dirac operator for a $K$-cycle.

Earlier, the $K$-homology of topological spaces had been developed as a functorial
theory whose cycles pair with vector bundles in the same way that currents pair with differential forms in the de Rham theory.   However,
the index theorem shows that the right partners for vector bundles are elliptic pseudodifferential operators (with the
pairing given by the index map), and Atiyah \cite{14} sketched how $K$-cycles should be recast in terms of elliptic operators. Kasparov found the right equivalence relation for such cycles and, more importantly, showed that the correct abstraction
of ``elliptic operator" is the notion of a Fredholm module over an algebra\cite{275}.
A $K$-$cycle$  over a pre-$C^*$-algebra $\alg$ is nothing other than a pre-Fredholm module.
Unfortunately computations with commutators $[F,a]$ coming from Fredholm modules can be quite cumbersome.
An important simplification was introduced by Baaj and Julg\cite{18}, who observed that if $\dd$ satisfies following
conditions then $F'=\dd(1+\dd)^{-1/2}$ determines a pre-Fredholm module $(\alg,\Hs,F')$. They also showed that all
$K$-homology classes of $\alg$ arise in this way. This motivates the following definition.


The definition of noncommutative geometry is achieved by extending the
situation from commutative algebra $C^{\infty}(M)$ to a noncommutative algebra
$\alg$.
Forgetting the space $M$, we will determine the Dirac operator algebraically.
The general Dirac operator $\dd$ is chosen so that
it satisfies following algebraic conditions giving a Hilbert space $\Hs$ for $\alg$.

\begin{definition}[spectral triple]A spectral triple---also called an `unbounded $K$-cycle'---
 for an algebra $\alg$ is a triple $(\alg, \Hs, \dd)$, where $\Hs$ is a Hilbert space carrying a
representation of $\alg$ by bounded operators (that we shall write simply $\xi\mapsto a\xi$ for the operator representing
$a\in\alg$), and 
\begin{enumerate}
\item $\dd$ is self-adjoint.
\item $[\dd,a]$ is a bounded operator for all $a\in A$.
\item compact resolvent $\Leftrightarrow(\dd^2+1)^{-1/2}$ is a compact operator.
\end{enumerate}
for each $a\in\alg$.
\label{def:stp}
\end{definition}

\subsection{Wodzicki residue}
To build a definition of gravity action out of the generalized Dirac operator, we give a relation between the symbols of an inverse generalized Laplacian and some intrinsic geometric quantities. 
First we shall review briefly  a few basic properties of pseudodifferential operators (see Ref.\cite{HeatKernel}).

Pseudodifferential operators are useful tools, both in mathematics and in physics. They were crucial for the proof of the
uniqueness of the Cauchy problem\cite{pdo1} and also for the proof of the Atiyah-Singer index formula\cite{pdo2}.
In quantum field theory they appear in any analytical continuation process (as complex powers of differential operators, like
the Laplacian operator)\cite{Seeley}. 
They constitute nowadays the basic starting point of any rigorous formulation of quantum field theory.

In the following $M$ is a compact $n$-dimensional Riemannian manifold and $E$ and $E'$ are (complex) vector bundles of rank $r$ and $s$ on $M$.
A $d$th order differential operator $L:\Gamma(M,E)\mapsto \Gamma(M,E')$ acting on sections of $E$ may be written in local coordinates for suitable trivializations of $E$ as
\begin{equation}
(Lu(x))_i=\sum_{j=1}^r\sum_{|\alpha|=0}^d(-i)^{|\alpha|}a_{\alpha}^{ij}(x)\partial_x^{\alpha}u_j(x)\ \ \ \forall i=1,...,s,
\end{equation}
where $\alpha=(\alpha_1,...,\alpha_n),\ \alpha_i\in {\cal N}_0$ is a multi-index with $|\alpha|=\sum_{i=1}^n\alpha_i$, $\alpha^{ij}$ is an $r\times s$-matrix and $\partial_x^{\alpha}:=\partial^{\alpha_1}/\partial x_1^{\alpha_1}\cdot\cdot\cdot\partial^{\alpha_n}/\partial x_n^{\alpha_n}$.
Motivated by the Fourier representation (on $\RR^n$),
\begin{align}
f&=\frac{1}{(2\pi)^n}\int \hat{f}e^{ix\cdot \xi}d\xi\\
L(x,D)[f(x)]&=\frac{1}{(2\pi)^n}\int \sigma^L(x,\xi)\hat{f}(\xi)e^{ix\cdot\xi} d\xi, \label{eqn:pdo}\\
\sigma^L(x,\xi)&=\sum_{k\leqq d}\sigma^L_k(x,\xi):=\sum_{|\alpha|\leqq d} a_{\alpha}(x)\xi^{\alpha}.
\label{eqn:ppdo}
\end{align}
$\sigma^L$ is a symbol of a pseudodifferential operator and  $\sigma_k$ is homogeneous of degree 
$k$ (that is, $f(t\xi)=t^k f(\xi)$ for all $t>0$). 
The leading term $\sigma_d^L$ is called the principal symbol of $L$. 
Then the  rule of product of the symbol is
\begin{align}
\sigma^{A\circ B}=\sum_{\alpha\geq 0}\frac{i^{|\alpha|}}{\alpha !}\partial_{\xi}^{\alpha}\sigma^A\partial_x^{\alpha}\sigma^B.
\label{eqn:psd}
\end{align}

Integration on ordinary manifolds may be recast into a noncommutative mold due to the existence of an important functional
on pseudodifferential operators, called the residue of Wodzicki\cite{490-491} who realized its role as the unique trace
(up to multiples) on the algebra of classical pseudodifferential operators.
Our discussion is restricted to the case of compact boundaryless manifolds; for an excellent introduction
to noncommutative residues on manifolds with boundary, see the paper by Schrohe\cite{420}.

\begin{definition}[Wodzicki]
Given a classical pseudodifferential operator $A$, there exists a 1-density on $M$, denoted $\wres A$, whose local
expression on any coordinate chart is
\begin{align}
\wres_x A=|dx^n|\int_{|\xi=1|} \sigma_{-n}^A(x,\xi)|d\xi^{n-1}|.
\end{align}
This is the Wodzicki residue density. By integrating this 1-density over $M$, we get the Wodzicki residue as noncommutative
residue functional,
\begin{align}
\Wres A:=\int_M \wres_x A.
\label{eqn:wzr}
\end{align}
\label{def:wzr}
\end{definition}

By this definition we have essential properties of the residue, $\Wres([A,B])=0$ and $\Wres(\sigma_k^A)=0$ for $k\neq n$ (see appendix \ref{app:wod}).
Furthermore, we regard the Wodzicki  residue is unique trace of pseudodifferential operator up to multiplication by constant.

\subsection{Dixmier trace and Connes' trace theorem}

An {\it infinitesimal} operator $T$ on an infinite-dimensional Hilbert space $\Hs$ with a countable orthonormal basis is
one for which, roughly speaking, $||T||<\epsilon$ for any $\epsilon>0$. Of course, the only operator satisfying this condition, as stated is $T=0$. If, however, we first shave off a finite-dimensional subspace of $\Hs$ before computing the norm of
$T$, then there is an ample supply of infinitesimals namely the space $\cop$ of compact operators on $\Hs$.

In noncommutative geometry, we require that such infinitesimals can  be related to the Lebesgue integration by any
linear map which is a kind of trace operator.
Using ordinary trace, an analogue of Lebesgue integration theory  was developed in the fifties by I.E.Segal\cite{427};
for instance, the Schatten ideal ${\cal L}^p$ is the analogue of the usual $L^p$ space\footnote{Several subclass of compact operators may be determined by imposing suitable conditions on the singular values. If $1\leqq p<\infty$, we may define the Schatten $p$-class ${\cal L}^p$ by $T\in {\cal L}^p \Leftrightarrow \sum_{k=0}^\infty \mu_k(|T|)^p<\infty$. This is ideal in $\cop$ and hence also in $\bop$. }. 
Later Dixmier \cite{136} discovered another kind of trace functional with a larger domain of compact operators, which turns out to be the right one to relate the trace and the integration  in noncommutative geometry.

The Dixmier trace $\tr^+$ is introduced as
\begin{equation}
\tr^+ T=\lim_{N\rightarrow\infty}\frac1{\log N}\sum_i^N\mu_i(|T|),\ \  |T|=(T^{\dagger}T)^{1/2},
\end{equation}
where $\mu_i$ is the decreasing series of eigenvalues.

Then we define the Dixmier ideal as
\begin{definition}
The Dixmier ideal of compact operators is defined by
\begin{align}
\Dc:=\left\{ T\in\cop:||T||_{1+}:=\sup_{N\geq N_0}\frac1{\log N}\sum_i~N\mu_i(|T|)<\infty\right\}
\end{align}
Clearly $||\cdot ||_{1+}$ is a norm. It can be easily seen that ${\cal L}^1\subset \Dc\subset {\cal L}^p$ for any $p>1$.
\label{def:did}
\end{definition}

It is achieved by the Connes' trace theorem to relate the Dixmier's trace functional and the noncommutative integration.
\begin{theorem}[Connes]
Let $H$ be an elliptic pseudodifferential operator of order $-n$ on a complex vector bundle $E$ on a compact Riemannian
manifold $M$. Then $H\in \Dc$, indeed $H$ is measurable, and 
\begin{equation}
\Dtr H=\frac{1}{n(2\pi)^n}\Wres H.
\end{equation}
\label{thm:con}
\end{theorem}
This Connes' trace theorem allows us to compute the integral of any function on a Riemannian manifold by an operatorial
formula. Thus, it is the starting point of a generalization of the notion of integral.
\begin{corollary}
For any function $a\in \Cinf(M)$,
\begin{equation}
\int_M a(x) dv=\frac{n(2\pi)^n}{\Omega_n}\Dtr (a\Delta^{-n/2}) .
\end{equation}
\end{corollary}
The compactness of $M$ is not requisite. Actually any integrable function $a$ on a Riemannian manifold $M$ satisfies
the above relation.


We recall that  an operator $T$ in Hilbert space is compact iff for any $\epsilon>0$ one has $||T||<\epsilon$ except on a finite dimensional subspace of $\Hs$. 
The size of compact operator $T$ is measured by the decreasing sequence $\mu_n$ of eigenvalue of $|T|=\sqrt{T^*T}$. The order of such an infinitesimal is measured by the rate at which these characteristic values $\mu_n(T)$ converge to $0$ when
$n\rightarrow \infty$.

One can show that all the conventional rules of calculus are valid, e.g. the order of $T_1+T_2$ or of $T_1T_2$ are as they should be ($\leq \alpha_1\vee\alpha_2$ and $\alpha_1\alpha_2$). Moreover the trace is logarithmically divergent for infinitesimals of order $1$, since $\mu_n(T)=O(\frac1{n})$.
Then the coefficient of the logarithmic divergence does yield an additive trace (just Dixmier trace) which in essence evaluates
the {\it classical part} of such infinitesimals. This trace vanishes on infinitesimals of order $\alpha >1$.


\subsection{gravitation action in noncommutative geometry}

Since as stated in the section \ref{subsec:dirac}, it is possible to give geometrical values by the Dirac operator $\dd$, we should show
\begin{align}
\Wres(|\dd|^{2-n})&=-\frac{2^{\lfloor n/2\rfloor}(n-2)\Omega_n}{24}\int_M R dv ,\nonumber\\
or \ \ \ \ \ 
\Wres(\Delta^{1-n/2})&=\frac{(n-2)\Omega_n}{12}\int_M R dv,
\label{eqn:gac}
\end{align}
in the following.
Of course, the Lichnerowicz formula tells that they are equivalent and
we can handle the both of them.
To show this relation, it is most simple to calculate $\sigma_{-n}^{\Delta^{1-n/2}}$.
Using Riemannian normal coordinate, one finds that \cite{CDW} from 
\begin{equation}
1=\sigma^{\Delta\circ \Delta^{-1}}=\sum_{\alpha\geq 0}\frac{i^{|\alpha|}}{\alpha !}\partial_{\xi}^{\alpha}\sigma^{\Delta}\partial_x^{\alpha}\sigma^{\Delta^{-1}}
\end{equation}
and its decomposition according to degrees.
Alternatively for our convenience we give a derivation from the asymptotic expansion of heat kernel and
the zeta-functional analysis (see appendix \ref{app:heat}).

The relation between the Wodzicki residue and the asymptotic expansion of heat kernel becomes
\begin{align}
\wres |\dd|^{k-n}=\frac{(2\pi)^nd}{\Gamma((k-n)/2)}a_k(x,x)|dx^n|,
\end{align}
where $a_k$ is given by Riemannian geometry.

Moreover from the well-known \cite{HeatKernel} fact $a_2(x;|\dd|^2)=-R(x)/3(4\pi)^{n/2}$ where $R$ denotes the scalar curvature of $M$, we can obtain the Kastler-Kalau-Walze formula\cite{CDW};
\begin{align}
\Wres |\dd|^{-n+2}=-\frac1{3}\int_M\Omega_n(\frac{n}{2}-1)R dv\ .
\end{align}

In this section, we have shown that the Wodzicki residue (which is the trace of pseudodifferential operator) of $\Delta^{-n/2+1}$
 or $\dd^{-n+2}$ provides the Einstein-Hilbert action.
From this and the fact that the geodesic distance of a Riemannian metric is given by the Dirac operator (the first of Connes' axioms), we are encouraged
to adopt a Dirac operator  as a dynamical variable of spacetime geometry. Indeed, we see that the variational principle of the Wodzicki-Einstein-Hilbert action for the Dirac operator leads to the Einstein equation.

In the philosophy of noncommutative geometry, the concept of spacetime manifold is forgotten and operator algebra is substitution of that.
There the Dirac operator is given as a spectral triple (Def. \ref{def:stp}) and called a quantum Dirac operator.
Then we need a candidate of the trace other than the Wodzicki residue to formulate the gravitational action for the quantum Dirac operator.
In the previous section, it was reviewed that Connes showed that the Dixmier trace (Def. \ref{def:did}) can play the role of the Wodzicki residue
in the Dixmier ideal (Def. \ref{def:did}).
Nevertheless, we should be noticed that $\Delta^{-n/2+1}$ and $\dd^{-n+2}$ are not included in the Dixmier ideal (see the reference\cite{ENG} or one can easily confirm that as an exercise) since the Dixmier trace of the homogeneous
 degree $\lambda>-n$ 
diverges while the Wodzicki residue of that vanishes. Then any regularization is required, which might be artificial.

In the present work, the regularization is related to the truncation of high frequency like the discritization of
 spacetime.
Considering a truncated Hilbert space, one make the Dixmier trace  related to a continuous limit
by the regularization (it is postponed  until section \ref{sec:via}).
In the case that the Dirac operator is to be classical as an pseudodifferential operator, the variational principle
determines only $\sigma_1^\dd$ and $\sigma_0^\dd$, while the other parts of different degree is determined to be zero,
 because $\Wres (\dd^{-n+2})$ only depends on $\sigma_{-n}^{\dd^{-n+2}}$.
On the other hand, for the quantum Dirac operator, we need to think over the EOM and Wodzicki-Einstein-Hilbert
action.
In the following sections we realize that, replacing the Dixmier trace by the ordinary trace by a truncated Hilbert space (e.g.,
in the discritization of the spacetime).
We call the gravitational system represented by the operator algebra `operator geometry'.

\section{canonical formulation of operator geometry}
To investigate the dynamics of operator geometry, in classical or quantum mechanically, we have to
 develop its analytical dynamics. Conventionally, analytical dynamics is described by flow and functions in the phase
space. An evolution of system is represented by a curve $q_i$ there, which maps $\RR$ into the configuration space $\RR^n$,
$q_i:\RR\mapsto \RR^n\ni q_i(t)$. 

On the algebraic abstraction of this commutative analytical dynamics, the mapping $q_i$ is replaced by a multiplet of operator
$Q_i\in \lop$ which act on a Hilbert space $\Hs$. Formally the algebraic counterpart of the mapping $q_i$ is given by a
quadratic form $\widehat{Q_i}(\phi)$ associated to $Q_i$
\[\widehat{Q_i}:\Hs\mapsto
\RR^n,\ \phi\in \Hs,\ <\phi|Q_i|\phi>\in \RR^n, i=1,...,n.\]
where the state vector $\phi$ varies in place of time parameter $t$.
Application to operator geometry will be straightforward. Motivated by the feet that the standard Einstein
Hilbert action is reproduced via
eq.(\ref{eqn:gac}), we take the Dirac operator as a dynamical variables.

First  we establish the abstraction of the ordinary analytical dynamics and then we will apply it to 
noncommutative gravity of the operator geometry.
\subsection{algebraic abstraction of analytical dynamics}
Now we give a natural algebraic abstraction of ordinary analytical dynamics by explicitly showing 
 no less than two general operators and its representation in a dynamical system.

In ordinary analytical dynamics, a configuration variable is a mapping $q_i:\RR\mapsto \RR^n\sim M$.
Algebraically, this classical variables $q_i$ and its derivative $\dot{q}_i$ with respect to time parameter $t$
  form a commutative algebra. 
Then they are reformulated by a representation of multiplication operators $M_q, M_{\dot{q}}\in \lop$, 
 the bounded operator on a Hilbert space $\Hs$.
Here the differential operator $D$ of $d/dt$ is introduced as a noncommutative operator so that $M_{df/dt}$ is given
by $[D,M_f]$ with $[M_f,M_{df/dt}]=0$ in the case of ordinary commutative dynamics. We
 may extend $D$ to a generalized noncommutative operator and forget its classical origin.
In our algebraic formulation of dynamics the evolution of the system is generated by the differential
 operator $D$ instead of 
the derivative with respect to time $t$.
We will forget the time parameter $t$ from now on.

In the following $A,B,D,Q$ are bounded operators on a Hilbert space $\Hs$. 
On the Hilbert space, the action integral is written as a generalized trace $\tr$.
Here `generalized' means that the $\tr$ is not required to be the ordinary trace but any positive linear functional $\tr: \lop\mapsto \CC$ with $\tr (AB)=\tr (BA)$.
 The explicit definition of the $\tr$ will be determined depending on the summabilities of concerned operators. For instance, we often consider the Wodzicki residue $\Wres$ as a trace for pseudodifferential operators or the
Dixmier trace $\Dtr$ for compact operators not only in the trace class, on an infinite dimensional $\Hs$,
 instead of the ordinary trace.

With $\tr [A,B]=0$\,  one can always cyclically change the order of operators in this trace.
Since the trace corresponds to noncommutative integral, this will give an analogue of integration by parts
 because 
\begin{align}
\tr (A[D,B])&=\tr (ADB-ABD)\\
&=\tr (ADB-DAB)\\
&=-\tr ([D,A]B).
\label{eqn:iip}
\end{align}

Though in the ordinary commutative case an operator $Q$ and its commutator $[D,Q]$ are supposed to be commutative with
each other, we do not assume this in our algebraic formulation since we think of quantum
noncommutativity as an eventual target.
We will give a Lagrangian naturally considering $Q$ to be a variable and $[D,Q]$ its time derivative.

In the conventional analytical dynamics on $M$,  an orbit is a differentiable mapping $(q_i,\dot{q_i}):\RR\mapsto \tm$ 
and the Lagrangian is defined as a function on $\tm$.
In contrast in the dynamics of operators, the variable operators $Q_i$ gives an orbit as a quadratic 
form $(\widehat{Q}_i,\widehat{[D,{Q}_i]})(\phi)$ on $\Hs$ and an one-parameter family $\phi(s)$ of the state vector $\phi\in\Hs$,
\[
(\widehat{Q}_i,\widehat{[D,Q_i]}):\Hs\mapsto \tm,\ \  \phi\in \Hs,\ \  (<\phi|Q_i|\phi>,<\phi|[D,Q_i]|\phi>)\in \tm .
\]
The Lagrangian should  be a mapping from $C^\infty(\RR,\tm)$ to $C^\infty(\RR)$ 
rather than a function of $\tm$, so as to reflect the noncommutativity of the operators.
Since $(<\phi|Q_i|\phi>,<\phi|[D,Q_i]|\phi>)$ cannot be reduced  to $L(<\phi|Q_i|\phi>,<\phi|[D,Q_i]|\phi>)$ in general, 
the Lagrangian
should be defined as $L(Q_i,[D,Q_i])\in \lop$ on $(Q_i,[D,Q_i])\in \lop^{2n}$,

Now we give an action of a simple one-variable Lagrangian, $L(Q,[D,Q])\in\lop$.
 supposing that $Q$ is the only variational operator.
Using the integration by parts eq.(\ref{eqn:iip}), the variation of the action is given by
\begin{align}
S&=\tr (L(Q,[D,Q]))\\
\delta S&=\tr \left(\frac{\partial L}{\partial Q}\delta Q+\frac{\partial L}{\partial([D,Q])}\delta([D,Q])\right) \\
&=\tr \left(\frac{\partial L}{\partial Q}\delta Q-[D,\frac{\partial L}{\partial([D,Q])}]\delta Q\right),
\end{align}
where the operator derivative $\partial/\partial Q$ is defined in appendix \ref{app:oar}.
Note that although the definition of the derivative
\begin{align}
\delta F=\frac{\partial F}{\partial Q}\delta Q
\end{align}
is not unique because of the operator ordering, it is unique inside the  trace.

Once the variational principle $\delta S=0$ is applied, an analogue of an usual Euler-Lagrange equation is obtained as
\begin{align}
\frac{\partial L}{\partial Q}-[D,\frac{\partial L}{\partial([D,Q])}]=0.
\label{eqn:ele}
\end{align}

Here we are required to be careful about the definition of the trace.
For the ordinary definition of the trace, $\tr (F\delta Q)$ implies, $F=0$ if $\delta Q$ is arbitrary. Moreover even if $\delta Q$ is restricted to be self adjoint, it concludes $F=0$ as long as $F$ is also self adjoint. 
In the present formulation $F$ must be self-adjoint since $L$ can be rewritten in self adjoint form under 
the rearrangement of the operator ordering inside the trace.
 
Nevertheless another definition of the trace would be adopted on an infinite dimensional Hilbert space, and the variational
principle derives other equations.
Though in the next section we actually treat such a case, in the rest of this section we proceed adopting the ordinary 
definition
of the trace.

A momentum operator conjugate to $Q$ is given by
\begin{align}
P&=\frac{\partial L}{\partial([D,Q])}.
\label{eqn:cmm}
\end{align}
In the present formulation we do not suppose that $P$ is  a multiplicative and commutes with $Q$ as in the ordinary commutative case.

Then Hamiltonian $H$ is defined by a Legendre transformation; 
\begin{align}
 H(P,Q)&=P[D,Q]-L(Q,[D,Q])\\
\delta H&=\frac{\partial H}{\partial Q}\delta Q+\frac{\partial H}{\partial P}\delta P \\
&=[D,Q]\delta P+P\delta([D,Q])-\frac{\partial L}{\partial Q}\delta Q-\frac{\partial L}{\partial([D,Q])}\delta([D,Q]).
\end{align}
Sometimes the first term of $H(P,Q)$ might be reordered in the trace so that $H$ is also a self adjoint operator.
From the definition of the conjugate momentum eq.(\ref{eqn:cmm}), the terms proportional to $\delta([D,Q])$  vanish. Comparing remaining proportional to $\delta Q$ and the one to $\delta P$,
the Hamilton equations are given by
\begin{align}
[D,Q]&=\frac{\partial H}{\partial P}\\
[D,P]&=-\frac{\partial H}{\partial Q},
\end{align}
where the Euler-Lagrange equation eq.(\ref{eqn:ele}) is used to derive the second equation.

Here one should be careful that $H$ is not a generator of evolution in the sense of the commutator but of  Poisson bracket,
while $D$ is the generator of evolution for the commutator 
\begin{align}
<\phi|[D,X]|\phi>=\{X, H\}_{\phi}.
\end{align}
Intuitive definition of the Poisson bracket $\{A,B\}_{Q,P}=\partial_Q A\partial_P B-\partial_P A\partial_Q B$ 
tends to be inconsistent because of the operator ordering eq.(\ref{eqn:lrd}) in appendix\ref{app:oar}.
The Poisson bracket is with complicated nature in noncommutative operator dynamics because of the operator ordering.
The precise definition of the Poisson bracket will be given later on.

Of course, the present formulation includes the ordinary analytical dynamics.
Usual dynamical variable $q(t)$ on time interval $[0,1]$ can be viewed as a multiplicative operator $M_q:f\mapsto qf, f\in L^2([0,1])$, while
$D$ is a representation of $d/dt$ on the Hilbert space $L^2([0,1])$.

In the case that $D$ is also a variational operator, we proceed
\begin{align}
S&=\tr (L(Q,[D,Q],D))\\
\delta S&=\tr\left( \frac{\partial L}{\partial Q}\delta Q+\frac{\partial L}{\partial([D,Q])}\delta([D,Q])+\frac{\partial L}{\partial D}\delta D \right)\\
&=\tr \left(\frac{\partial L}{\partial Q}\delta Q-[D,\frac{\partial L}{\partial([D,Q])}]\delta Q+[Q,\frac{\partial L}{\partial [D,Q]}]\delta D+\frac{\partial L}{\partial D}\delta D\right),
\end{align}
where the last line follows from a relation
\begin{align}
\tr\left(A\delta([D,Q])\right)=\tr\left(A[\delta D,Q]+A[D,\delta Q]\right)=\tr\left([Q,A]\delta D-[D,A]\delta Q\right).
\end{align}

The variational principle for each variable gives
\begin{align} 
-[D,\frac{\partial L}{\partial [D,Q]}]+\frac{\partial L}{\partial Q}&=0\\
 [Q,\frac{\partial L}{\partial [D,Q]}]+\frac{\partial L}{\partial D}&=0.
\label{eqn:cre}
\end{align}

The second equation is not a dynamical equation of $D$ but a constraint equation in commutative sense, 
since there is no kinetic term of $D$.
For a dynamical variable $Q$, a conjugate momentum is given by $\partial L/\partial [D,Q]$.
Then a Hamiltonian associated with a constraint equation (\ref{eqn:cre})is given by a Legendre-transformation;
\begin{align}
H(P,Q,D)&=P[D,Q]-L(Q,[D,Q],D).
\label{eqn:crh}
\end{align}

In the following case of gravity, the constraint equation(\ref{eqn:cre}) can be algebraically solved as a function of $P$
and $Q$; $D=D(P,Q)$ 
since the equation explicitly depends on $D$.
Then we can substitute it to the constrained Hamiltonian eq.(\ref{eqn:crh}).
Finally we will get a Hamiltonian $H(P,Q,D(P,Q))$ without any constraint.

\subsection{algebraic abstraction of Einstein-Hilbert action}

Now we develop the algebraic canonical formalism of gravity in the context of operator geometry.
For   convenience and usefulness, we adopt the Dirac operator rather than the Laplace operator as a dynamical
 variable operator.
Then in the following, we regard the operator as a superrepresentation of  Clifford algebra on a complex superspace
$E=E^+\oplus E^-$ that respects $\ZZ_2$-grading\cite{282}. A superalgebra is a superspace with product $A^i\cdot A^j=A^{i+j} (i,j\in {\bf Z})$.

In this case, we naturally generalize the commutator to a supercommutator
\begin{align} 
\scom{a,b}=ab-(-1)^{|a||b|}ba
\end{align}
$|a|=0$ for $a\in A^+$ and $|a|=1$ for $a\in A^-$.

A trace on the superrepresentation is naturally defined as a supertrace:
\begin{align}
\str(a)= \begin{cases}\tr_{E^+}(a)-\tr_{E^-}(a) & a \ {\rm even} \cr
0 & a \ {\rm odd}
\end{cases},
\end{align}
where $\tr_{E^{\pm}}$ is the partial trace on $E^{\pm}$.
Of course, the previous requirement to the trace is satisfied for $\str(\scom{A,B})=0$.
 
It can be easily confirmed that the integration by parts of super algebra is
\begin{align}
\str (A\scom{D,B})=\str(\scom{A,D}B)=\str( -(-1)^{|A||D|}\scom{D,A}B).
\end{align}

In the following we consider the case of $n=\dim(M)=4$.
To give a canonical formalism, we suppose that spacetime $M$ is
topologically $\Sigma\times I$, so that $C(M)$ is isomorphic to $C(\Sigma)\otimes C(I)$.  
On $M$,
the Dirac operator is then decomposed as 
\begin{align}
\dd&=i\gamma^\mu\nabla_\mu=D+D_t, \\
D&=i\gamma^i(t,x)(\partial_i+\omega_i(t,x)),\ \  \gamma^i=\gamma^Ae_A^i(t,x),\\
D_t&=i\gamma^t(t,x)(\partial_t+\omega_t(t,x)),\ \ \gamma^t=\gamma^Ae_A^t(t,x),
\end{align}
$e_A^{i\ or\ t}$ is tetrad, $\gamma^A$ is a representation of the Clifford algebra, $\nabla_\mu$ is 
the covariant derivative and $\omega_\mu$ is the coefficients of spin connection.
The general form of $\dd^2$ is given by
\begin{equation}
\dd^2=D^2+\{D_t,D\}+D_t^2=D^2+\scom{D_t,D}+D_t^2,
\end{equation}
where $D$ and $D_t$ are of $A^-$.

To understand the kinematical role of $\scom{D,D_t}$, we write down that in the case of ordinary spacetime geometry,
with  the Clifford algebra $\gamma_\mu\gamma_\nu+\gamma_\nu\gamma_\mu=g_{\mu\nu}$,
\begin{align}
\scom{D_t,D}&=-\scom{\gamma^t\nabla_t,\gamma^i\nabla_i}=-\gamma^t\nabla_t
\gamma^i\nabla_i-\gamma^i\nabla_i\gamma^t\nabla_t\\
&=-\gamma^t[\nabla_t,\gamma^i\nabla_i]-\gamma^i[\nabla_i,\gamma^t\nabla_t]+\gamma_t\gamma_i[\nabla_t,\nabla_i]\\
&=:D_t(D)+D(D_t)+\gamma_t\gamma_i[\nabla_t,\nabla_i].
\label{eqn:scom}
\end{align}

From the definition of $D_t$, we think of $D_t$ as the operator of evolution in place of $\partial /\partial t$.
We regard that $D$ is a dynamical variable,
while $D_t$ is not because $\dd^2$ does not contain $[\nabla_t,\gamma^t\nabla_t]$ even in $D_t^2$. 
Since $D_t(D)$ is included in the eq.(\ref{eqn:scom}),
$\scom{D_t,D}$ will be considered as the time evolution of $D$.

Considering four dimensions and omitting the gravitational constant, from the Connes axioms the action is given by 
the inverse square of the Dirac operator;
\begin{align}
S&=\tr\dd^{-2}=\str \chi\dd^{-2}=\str\left( \frac{\chi}{D^2+\{D_t,D\}+D_t^2}\right)\\
&=\str\left( \frac{\chi}{D^2+\scom{D,D_t}+D_t^2}\right),
\end{align}
where $\chi$ is the $\ZZ_2$-grading operator defined by $\chi(v)=(-1)^{|v|}v$ when $v$ is either even or odd
and the supertrace is given by $\str T=\tr(\chi T)$.

With $\G(\dd)=\chi\dd^{-4}$, the variation of action is
\begin{align}
\delta S&=\str\left(-\G(\dd)(D\delta D+\delta D D+\delta\scom{D,D_t}+D_t\delta D_t+\delta D_t D_t)\right)\\
&=\str((-\G(\dd)D+D\G(\dd)-\scom{D_t,\G(\dd)})\delta D \nonumber\\
&+(-\G(\dd)D_t+D_t\G(\dd)+\scom{D,\G(\dd)})\delta D_t),
\label{eqn:ove}
\end{align}
where we use $\str\scom{\ ,\ }=0$ instead of $\tr [\ ,\ ]=0$ in order to arrange the order of operators.

Since the $D_t$ is not a dynamical variable, the last term gives a constraint equation.
Comparing with the ordinary Einstein equation in the first order formalism (that is, for example, the Einstein-Cartan formalism), we can see
the fact that both non-dynamical variables $e^t$ and $\omega^t$ in the ordinary
 formalism are contained only in the non-dynamical variable $D_t$ in the present formalism.

Taking the advantage of one of the virtues of our formulation that the constraint equation can be formally solved, we will go over to the Hamiltonian formalism. The conjugate momentum of $D$ is given by
$\partial L/\partial\scom{D_t,D}$;
\begin{equation}
P_D=-\frac{\chi}{(D^2+\scom{D_t,D}+D_t^2)^{2}}.
\label{eqn:cmom}
\end{equation}

The Hamiltonian is given by
\begin{align}
H&=P_D \scom{D_t,D}-L\\
&=\frac{\chi}{(D^2+\scom{D,D_t}+D_t^2)^{2}}(-2\scom{D_t,D}-D^2-D_t^2)\\
&=-2\chi(-\chi P_D)^{1/2}-P_D(D^2+D_t^2),\label{eqn:ham}
\end{align}
where the substitution $\scom{D_t,D}=(-\chi P_D)^{-1/2}-D^2-D_t^2$ is made using the definition of conjugate momentum eq.(\ref{eqn:cmom}).
The variational principle will give the Hamilton's equation as we see below. Using $\str{\ ,\ }=0$, the variation of the action is given by
\begin{align}
S&=\str \left(P_D\scom{D_t,D}-H(P_D,D,D_t)\right)\\
\delta S=&\str (\scom{D_t,D}-(-\chi P_D)^{-1/2}-D^2-D_t^2)\delta P_D\nonumber\\
&+(-\scom{D_t,P_D}+P_DD-DP_D)\delta D \nonumber\\
&+(-\scom{D,P_D}+P_DD_t-D_tP_D)\delta D_t.
\label{eqn:heq}
\end{align}

Here we have assumed the Hilbert space on which the operator polynomial of quantum Dirac operator $\dd$
is traceclass and the trace is given by the ordinary trace, for a while.
This implies that a quantum space is, in some sense, `discretized'.
If the operator polynomial in eq. (\ref{eqn:heq}) were a traceclass operator and the definition of the trace were the ordinary trace, the equation would simply imply the operator equations (\ref{eqn:ev1})(\ref{eqn:ev2}).
Since the operator polynomial is actually neither traceclass nor in the Dixmier trace class $\Dc$, however, for the classical Dirac operator of first order pseudodifferential operator and its trace is the Wodzicki residue,
 we will derive valid EOMs for the classical Dirac operator via a rather complicated process in the next
section. The EOMs will be identified to the Einstein equation.

Though as mentioned in the definition of the spectral triple (Def.\ref{def:stp})
 $\dd$ and $\delta \dd$ is restricted to be self-adjoint, the variational principle concludes the
following equations under the calculus of supertrace in the Clifford algebra;
\begin{align}
&\scom{D_t,D}-(-\chi P_D)^{-1/2}-D^2-D_t^2=0\label{eqn:ev1}\\
&-\scom{D_t,P_D}-\scom{D,P_D}=0\label{eqn:ev2}.
\end{align}

Though the $\delta D_t$ term in eq.(\ref{eqn:heq}) gave the constraint equation, the constraint equation and the evolution
equation are degenerated in this formulation.
It might be thought to be strange that there is no constraint equation to be imposed on the initial data. 
In this algebraic formulation, however, the spacetime manifold is forgotten and the equation is not only on the initial 
Cauchy surface but  on the whole spacetime.
Usually the constraint equation at the initial surface is assured on the whole spacetime by the evolution equation.
So, in this case the equation (\ref{eqn:ev2}) serves both as a evolution equation and a constraint equation.
In other words, though for a initial condition of the evolution equation the constraint equation will be required,
in this algebraic formulation the initial condition is realized as taken in as the arbitrariness of the algebraic equation.
This means the evolution equation and the constraint equation are unified into the equation(\ref{eqn:ev2}) by the present
algebraic treatment.

If in a quantum system, we are allowed to take this regularization of the trace,
we can discribe quantum gravity by the Hamiltonian (\ref{eqn:ham}), where $D$ and $P_D$ are a dynamical variable and its conjugate
momentum and time derivative operator $D_t$ is arbitrary.

\section{gravitation via operator geometry}\label{sec:via}
The main purpose  of the present article is to reformulate a canonical formulation of the Einstein gravity on the base of 
operator analysis of pseudodifferential operator
on $C^\infty(M)$ and to extend it to quantum gravity in noncommutative space.

We have completed the former in the previous sections. The latter will be achieved by replacing $C^\infty(M)$ for an arbitrary algebra $\alg$, which is not required to be commutative.
In this section we will give a generalized Dirac operator to represent geometry on $\alg$.
If $\alg$ is a noncommutative deformation of any commutative algebra, like $C^{\infty}(M)$,
 with conventional representation and a Dirac operator, 
we can obtain the generalized Dirac operator in a straightforward way by introducing the noncommutative product
into the definition of the Dirac operator. For instance, with $[x_\mu,x_\nu]=i\theta_{\mu\nu}$ the ordinary Dirac operator
is the right one since $i\gamma^\mu(\partial_\mu+\omega_{\mu})$ inherits the noncommutativity.

If we consider $\alg$ to be a general $C^*$-algebra, an algebraic definition of the Dirac
operator will be required.
It was already given as the definition of the spectral triple (Def.\ref{def:stp}).
From now on we call it  quantum Dirac operator to distinguish it from the ordinary (classical) Dirac operator $\dirac$.
In the following, we discuss the relation between the operator geometry in the previous section, and  classical and
quantum gravity.

\subsection{regularization and correction}

First we consider the $n$-dimensional case of $\alg=C^\infty(M)$ and $\dd$ to be the ordinary Dirac operator on it. For a pseudodifferential operator as the classical
Dirac operator, we demonstrate that the Einstein equation is reproduced from the operator
variational equation (\ref{eqn:ove}).
Even if we get a variational equation
\begin{align}
\delta S=\tr\left[\sum_iF_i({\bf Q})\circ\delta Q_i\right]=0,
\end{align}
it does not necessarily mean that the EOM is given by $F_i=0$ for general definition of trace.
Here the trace is equivalent to Wodzicki residue, except the case of homogeneous degree $k\neq -n$,
\begin{align}
\tr (F_i\circ \delta Q_i)&=\Wres (F_i\circ \delta Q_i) \\
&=\Wres (\sigma_{-n}^{F_i\circ \delta Q_i}). 
\end{align}
To derive EOMs, the rule of product in eq.(\ref{eqn:psd}) should be applied as
\begin{align}
\sigma_{-n}^{F_i\circ \delta Q_i}=\sum_\beta\sum_{\alpha\geq 0}\frac{i^{|\alpha|}}{\alpha !}\partial_\xi^\alpha\sigma_{-n+\alpha-\beta}^{F_i}
\partial_x^\alpha\sigma_\beta^{\delta Q_i}.
\end{align}

In our case, since the variational variable $Q_i$ is a Dirac operator $\dd$,
\begin{align}
\dd&=\dirac=i\gamma^\mu(\partial_\mu+\omega_\mu),\\
 \sigma_1^\dd&=\gamma^\mu\xi_\mu,\ \ \sigma_1^\dd=i\gamma^\mu\omega_\mu,
\label{eqn:ddd}
\end{align}
 its variation should be restricted to the homogeneous of degree $k=1$ or $0$.
Similarly it is easily found that the pseudodifferential operator of `equation of gravity' $F({\bf Q})=G(\dd)$ is of order $1-n$ since $G\circ\delta \dd$
is of order $2-n$ (eq.(\ref{eqn:gac})).
Then the contribution to $\Wres$ is
\begin{align}
&\sigma_{-n}^{G\circ \delta \dd}=\sigma_{-n}^{G}\sigma_0^{\delta \dd}+\sigma_{-n-1}^G\sigma_1^{\delta \dd}\\
&+i\partial_\xi\sigma_{-n+1}^G\partial_x\sigma_0^{\delta \dd}+i\partial_\xi\sigma_{-n+1-1}^{G}\partial_x\sigma_1^{\delta \dd}.
\end{align} 

Since $\sigma_0^{\delta \dd}$ and $\sigma_1^{\delta \dd}$ are arbitrary, integrating it by part, we get two equations for the variation 
of $\dd$;
\begin{align}
&\sigma_{-n}^G-i\partial_\xi\partial_x\sigma_{-n+1}^G=0 \\
&\sigma_{-n-1}^G-i\partial_\xi\partial_x\sigma_{-n}^G=0.
\end{align}
Finally we have 
\begin{align}
\sigma_{-n-1}^G-\partial_\xi^2\partial_x^2\sigma_{-n+1}^G=0.
\end{align}

From eq.(\ref{eqn:gac}) the equations are equivalent to the Einstein equation  
without matter under the torsion free condition.
The calculation is just equivalent to the first order formalism of the Cartan-Einstein-Hilbert action $\sim\int R(e,\omega) dv$.
Indeed, Connes showed the variational aspects (the third axiom of Connes') of $\Wres(\dd^{-n+2})$ and proved that the residue is minimum
at the Levi-Civita connection.
This agrees with the fact that the variational principle for $\delta \omega$ implies the torsion free
condition for the tetrad $e$.

On the other hand, for a quantum Dirac operator whose trace should be given not by the 
Wodzicki residue but by an algebraic trace i.e., the Dixmier trace, it is not straightforward to show how to get a corresponding classical EOM from
the variational principle. 
To achieve that, we have to cope with the problem that the Connes' trace theorem\cite{88} holds only for the operator in 
the Dixmier ideal, while the gravitational Lagrangian $|\dd|^{2-n}$ is not in the ideal.
This fact means the quantum Dirac operator would be related to a classical EOM through any regularization to subtract such
a quantum divergence.

To find a $\log N$ divergence part, which corresponds to the Dixmier trace, 
we would subtract background action from the action trace;
\begin{align}
S_{reg}&=S(\dd)-S(\dd_0)\\
&=\Dtr(L(\dd)-L(\dd_0)),
\end{align}
where $\dd_0$ is any background Dirac operator such that the divergence of $S$ is cancelled and $L(\dd)-L(\dd_0)$
 becomes a Dixmier class operator;
\begin{align}
\sigma_i(L(\dd))=\sigma_i(L(\dd_0)),\ \ \ \ i=2-n,1-n.
\end{align}
Then $S_{reg}$ is  related to $\Wres$, whose variational principle will give the Einstein-Cartan equations similar to
the classical Dirac operator.

Now we pay attention to quantum feature of the system.
While in quantization of ordinary gravity, problem is that general relativity is totally constrained system which
are highly nontrivial to solve, an attractive feature of the present formulation is that the constraint equation can be algebraically solved if we adopt the 
 ordinary trace definition.
Then we expect that the quantum physics is rather determined by the ordinary trace not by the Dixmier trace though the classical action of gravitation is related to the Dirac operator by the Dixmier trace.
Indeed while in quantum gravity the short wave-length divergence is fundamental difficulty, in our algebraic approach a
reinterpretation of the trace controls the divergence by like discritization
of spacetime which gives the high frequency cutoff.
Here it seems natural that a mathematically simple model by the ordinary trace represents quantum theory
and its logarithmic divergence of the Dixmier trace yields classical observables.
For that purpose, we consider a large but finite dimensional Hilbert space as a subspace of the original infinite dimensional Hilbert space.
For the finite dimensional Hilbert space, the definition of ordinary trace becomes the unique definition of trace functional.
Our regularization scheme for a quantum Dirac operator is summarized as the redefinition of the trace;
\[
\tr \rightarrow \Dtr \rightarrow \Wres.
\]

In other words, we can take $N$ large and define a truncated Dixmier trace by
\begin{align}
\tr_N T=\sum^N\frac1{\log N}\mu_i(|T|),\ \  |T|=(T^{\dagger}T)^{1/2}.
\end{align}
By this truncation, it is expected that $S_N=\tr_N (\dd^{-2}-\dd_0^{-2})$ shifts from the Wodzicki residue of $\dd^{-2}$
by quantum correction.
In fact, the additivity of the Dixmier trace is only approximately correct\cite{113} on finite dimensional Hilbert space.
That implies that $\tr_N[A,B]=0$ does not follow from $\tr_N AB=\tr_N BA$.
In the proof of Theorem \ref{thm:wzr}, the vanishing of trace of a commutator  leads to the result that
the Wodzicki residue depends only on the homogeneous component of degree $-n$.
So the correction will include the homogeneity of degree $k\neq -n$.

From eq.(\ref{eqn:hom}) and eq.(\ref{eqn:poisson}) in appendix \ref{app:wod}, we get
\begin{align}
f_k&=\frac{1}{n+k}\sum_{i=1}^n\frac{\partial(\xi_i f_k)}{\partial \xi_i}\\
&=\frac{1}{n+k}\sum_{j=1}^n-i[x^j,\xi_j f_k],
\end{align}
for a homogeneous term $f_k$ of degree $k$.

For positive Dixmier class operators $a,b$,
the additivity of $\tr_N$ approximately holds for large $N$ as \cite{113}
\begin{align}
\tr_Na+\tr_Nb-\tr_N(a+b)\sim \frac{\log\log N}{\log N}(||a||_{1+}+||b||_{1+})\rightarrow 0.
\end{align}

Then a rough estimation gives\footnote{It is not true that each traced operator is positive. Furthermore it is difficult to estimate the norm $||\cdot||_{1+}$ correctly.}
\begin{align}
 \tr_N f_k\sim O(\frac{\log\log N}{\log N})||f_k||
\end{align}

So if the correction of homogeneous with degree $k\neq -n$ exists, its magnitude is expected to be
\begin{align}
\delta S_k\sim \tr_Nf_k\sim \frac{\log\log N}{\log N}||f_k||.
\label{eqn:Ds1}
\end{align}

To evaluate the correction more precisely, we recall the Connes' trace theorem\cite{88}.
The asymptotic behavior of harmonic series is
\begin{align}
\sum^N\frac1{n}=\log N+\gamma+o(\cdot)
\end{align}
where $\gamma$ is the Euler's constant.
On the other hand, since the Euler's constant is also one of the coefficients of the Laurant expansion of $\zeta(s)$,
\begin{align}
\zeta(s)=\frac{1}{s-1}+\tau-\tau_1(s-1)+\tau_2(s-1)^2+\dots
\end{align}
where $\tau_i$ is the Stieltjes constants,
 we expect to evaluate the correction ($\tr^+A^{-1}-\tr_NA^{-1}$) near the pole $s=s_k$ of $\zeta_{A}(s)$.

The correction of classical action given by
$\Dtr(\dd^{-2})-\tr_N(\dd^{-2})$ is evaluated as 
\begin{align}
\delta S=[\Dtr(\dd^{-2})-\tr_N(\dd^{-2})]=\lim_{s\rightarrow 1}(\zeta_{\dd^2}(s)-\zeta_{\dd^2}(s+\epsilon(N))).
\end{align}

The relation between the correction and the Seeley-De Witt coefficients for finite $\epsilon$ in eq.(\ref{eqn:43}) in appendix \ref{app:heat}is given by
\begin{align}
&\frac1{\Gamma(s)}\int_0^\infty \sum a_k t^{(k-k')/2+\epsilon}\frac{dt}{t} \ ,\\
&=\frac1{\Gamma(s)}\sum a_k\int_{-\infty}^\infty e^{(\delta k/2 +\epsilon) \log t} d(\log t)\ , \\
&=\frac1{\Gamma(s)}\sum a_k\frac{1}{\delta k/2+\epsilon}\lim_{T\rightarrow\infty}e^{\delta k T/2},
\end{align}
with $\delta k=k-k'$.
For large $T$, we need regularization to take coincident limit $K(x,x)=\lim_{y\rightarrow x}K(x,y)$.
The result is
\begin{align}
\delta S&\sim\sum f_k \int a_k(x,x) dv\\
&\propto\frac1{(4\pi)^{n/2}}\int dv \left(\frac1{288}R^2-\frac1{180}R_{\mu\nu}R^{\mu\nu}-\frac7{1440}R_{\mu\nu\rho\sigma}R^{\mu\nu\rho\sigma}+\frac1{120}\square R
...\right)
\end{align}
This is the same as the result already observed in the Heat kernel evaluation of quantum field theory\cite{CD}.

\subsection{gauge symmetry and noncommutativity}

Though in the above, a new formulation for classical and quantum gravity has been developed, the physical meaning of noncommutativity of the formulation has not yet completely clarified.
One may think of two possibilities for the physical origin of the noncommutativity in this formulation. One is that we consider the noncommutative
space as determined by noncommutative algebra not by commutative algebra like $C(M)$.
In this case, it would be far reaching to discuss quantum gravity in such a noncommutative space.
Even in classical physics, it would be ``academic" and of no physical relevance to study such a noncommutative space whose topological structure
is something pathological.
The other is to consider the quantum Dirac operator $\dd$ for which the classical condition that   $[\dd,a]$ commutes with $a$
is not imposed and noncommutativity appears in this sense. 
This implies that in our formulation the algebra of dynamics $\langle a, [\dd,a] \rangle_\CC$ provides the basis for
quantum theory unless we impose the classical condition on the Dirac operator.
To understand that further one may discuss the gauge symmetry of this formulation.
Since $\alg =C^\infty(M)$ is the commutative algebra of smooth functions on $M$ one easily see that the following defines a one-to-one correspondence between diffeomorphism $\varphi$ of $M$
 and the automorphism $\alpha\in {\rm Aut}(\alg)$ of the algebra $\alg$ (preserving the $*$ involution),
\begin{equation}
\alpha(f)(x)=f(\varphi^{-1}(x)),\ \ \forall x\in M,\ f\in C^{\infty}(M).
\end{equation}

In noncommutative case an algebra always has also inner automorphism given by
\begin{equation}
\alpha_u(f)=ufu^*,\ \ \  \forall f\in \alg
\end{equation}
for any element of the unitary group $\cal U$ of $\alg$,
\begin{equation}
{\cal U}=\{u\in \alg;uu^*=u^*u=1\}.
\end{equation}
The subgroup Int$(\alg)\subset {\rm Aut}(\alg)$ of inner automorphisms is a normal subgroup and it will provide us with our
group of internal gauge transformations. 
To understand the physical significance of the noncommutativity, one should make an answer to the question of what
the symmetry of this inner automorphism is.

In the context of spectral gravity\cite{SG}\cite{GUT} in which the standard model with gravitation is realized as a noncommutative
 geometry,  since inner automorphisms and
internal symmetries actually coincide and both groups will have to be lifted to the spinors, it is concluded that the action of inner automorphisms on the metric gives rise to internal fluctuations of the latter.
That replaces
$\dd$ by $\dd+A+JAJ^{-1}$  and gives exactly the gauge bosons of the standard model, with its symmetry breaking Higgs sector, when we apply it to the  finite geometry.
 The spectral action when restricted to the special metrics will give the interaction Lagrangian of the bosons. 
Thus the only distinction that remains between matter and gravity is the distinction Int$(\alg) \neq {\rm Aut}(\alg)$ and
will vanish for a number of highly noncommutative algebra.
Then it turns out that the noncommutativity is only caused by the matter field.

On the other hand, a pure gravitational theory like the present formulation suffers the noncommutativity 
because of the finite
dimensional Hilbert space through
the quantization or regularization including any truncation of short wave length.
For example, in a finite matrix algebra, a diagonal matrix $a$ generates a noncommutative subalgebra
with an off-diagonal matrix $D$, $\langle a,[D,a]\rangle_\CC\ ;\ \ a, D\in M_n$ while for an element $f$ of $C^\infty(M)$, a differential operator
$D=\partial_t$ generates a commutative algebra by $\langle f,\dot{f}\rangle_\CC$, which is equivalent to classical condition
that $[D,f]=\dot{f}$ is multiplicative operator as well as $f$.

The physical significance  of this noncommutativity is that the short wave-length truncation essentially brings  noncommutativity to the system
even if it is of a single particle.
Recalling the Gel'fand-Na\u{i}mark theorem (Thm. \ref{thm:gn1}) we see that a character $\mu=\hat{x}$ which is a unitary equivalent class of irreducible representation play a role of a point in space. 
With a noncommutative unitary element, however, the eigenstate of position $|\hat{x}>$ loses its special role as the representation of classical mechanics in the Hilbert space.
Instead of that, a pure state $|\phi>$ which would be an arbitrary linear combination of the position eigenstetes
will get an essential role as the elements of quantum space. 
Of course, it has been encouraging that the pure state $|\phi>$ takes the place
of a Gel'fand representation $\hat{x}(f)=f(x)$, in the standard prescription of the noncommutative geometry.
In a sense this is a virtue of the present formulation since the discritization of space will furnish more general meaning
 of continuous representation via linear combination of  positions.

In other words, in quantum theory, we do not choose any particular representation among the unitary
equivalence classes of the system. In macroscopic scale, however, quantum correlation loses by physical processes
like the collapse of the wavefunction. A particular representation is chosen among the unitary equivalence class, i.e. the symmetry is spontaneously broken so that
a classical space emerges.

\section{conclusion and discussions}

We have developed a new formulation of gravity (`operator geometry') which is expressed in the context of operator algebra.
The conventional geometrical quantities like curvature are encoded in the Dirac operator algebraically defined as a spectral triple in the operator geometry.
A newly established algebraic abstraction of analytical dynamics together with the Hamiltonian formalism gives the dynamics of the operator geometry. Remarkably it is found that the constraint equation and the evolution equation are formally
unified under the regularization by the reinterpretation of the trace functional.

We have argued that the classical equation which is the continuous limit of discrete system by the quantum assumption
and the semi-classical effect as a correction of the continuous (large dimensional Hilbert space) limit.
We have found the semi-classical action agrees with the known result of quantum field theory\cite{CD}.
To realize the discretized aspect of space, the noncommutativity of operator geometry arises.
From that one may understand that the eigenstate of position loses its privileged classical meaning at quantum scale.
Each pure state in quantum space is the replacement of a point in the ordinary space.
A full quantization  of the operator geometry is postponed to the forthcoming work\cite{FCW}.

In the present article, the semi-classical correction is evaluated not as the correction by the quantum fluctuation
of fields but as the noncommutative difference  caused by the discritization of space.
These entirely different effects yield the same result as their calculations are structurally similar.
This paradox will be resolved when we clarify the relation between the present algebraic formulation and quantum field theory.
In the present work we have laid the groundwork for algebraic quantum gravity.
The full quantization would make the beginning to discuss the paradox.
The Poisson bracket become significant when we go to canonical quantization\cite{FCW}.
To give consistent one, we prepare total Hilbert space $\Hs_T=\Hs_Q\otimes\Hs$ of representation space 
for quantized operators as the tensor product of quantum Hilbert
space $\Hs_Q$ and the Hilbert space $\Hs$ in the present article for noncommutativity.
The Poisson bracket should be defined so as to correspond to the commutator on
the quantum Hilbert space $\Hs_Q$. Then we should detach the Poisson bracket from the noncommutativity of space.
The canonical quantization will be developed in the forthcoming work.

Along with the present work we have some possibilities of noncommutative space to study.
For example, a commutative algebra on a quotient space has the same spectral space as that of another algebra by crossed product in the noncommutative geometry\cite{ENG}.
Sometimes the quotient space suffers  topological irregularity; a well known example is the Lebesgue non-measurability of a
irrational rotation ring, which is isomorphic to noncommutative torus.
Even with such an irregularity, the present formulation would work; at least in the above example,  noncommutative
geometry is well established\cite{390}.
Why our macroscopic world is commutative and Lebesgue measurable so that we can recognize our spacetime
dimensions? 
We will find reasonable explanation for this question  by studying
the dynamics of such noncommutative spaces.
Moreover we expect that the formulation is also applicable to other singular spacetime.
At first sight, this seems probable for the singularity allowing a self-adjoint extension\cite{sae} since we have a sound
Dirac operator there without knowing the topological structure of the singularity.
After we understand the problem algebraically, $K$-theory\cite{kth} may help to understand that in topologically.

To include matter field, some ideas have been proposed\cite{GUT}. The simplest step is to include the cosmological constant.
Since $V=\int dv\propto \tr^+|\dd|^{-n}$ from the Connes theorem, the contribution of cosmological constant would be
$\Lambda\tr |\dd|^{-n}$.
In order to have gauge boson 
we extend the geometry to a fiber bundle which is determined as an idempotent algebraically.
Furthermore considering discrete geometry the Higgs sector is able to be contained.
The interaction with fermions, which has the right hypercharge assignment, is obtained directly from $<\psi,\dd\psi>$ (with $\dd+A+JAJ^{-1}$ instead of $\dd$), and is thus also of spectral nature being invariant under the full unitary group of operators in the Hilbert space. 
Along that the standard model is proposed by \cite{SG}. 
Hybrid of this model and the present formulation may be possible.

\section*{Acknowledgement}
We are greatly thankful to Professor A. Hosoya for his useful comments and careful reading of the manuscript.

\appendix
\section{Wodzicki residue}
\label{app:wod}
In this appendix, we give a important fact about the construction of Wodzicki residue.
Most of proofs are left out and can be found in the textbook\cite{ENG}.

Suppose $f$ is homogeneous of degree $\lambda$, that is $f(t\xi)=t^\lambda f(\xi)$.
A direct calculation yields following proposition.
\begin{proposition}
For any function $a_{-n}(\xi)$ homogeneous of degree $-n$, the form $a_{-n}\sigma_\xi$ on $\RR^n\setminus \{0\}$ is closed.
\end{proposition}
$\sigma_\xi$ is the volume form on the unit sphere $\{\xi:|\xi|=1\}$.

We shall consider integrals of the form $\int_{|\xi|=1}a_{-n}(\xi)|\sigma_\xi|$. From the above proposition, it follows that
they can be calculated using any section of the $\RR^+$-principal bundle $\RR^n\setminus \{0\}$.

On the other hand, homogeneous functions are generically sums of derivatives, on account of Euler's theorem: if $f$ is homogeneous of degree $\lambda$,
\begin{align}
\frac{1}{n+\lambda}\sum_{i=1}^n\frac{\partial(\xi_i f)}{\partial \xi_i}=\frac{1}{n+\lambda}(nf+\sum_j\xi_j\frac{\partial f}{\partial \xi_j})=f
\label{eqn:hom}
\end{align}
This arguments fails, however, for $\lambda=-n$. Instead, we get a more restricted result, from the fact that $a_{-n}\sigma_\xi$ is closed form.

\begin{lemma}
$\int_{|\xi|=1}a_{-n}(\xi)|\sigma_\xi|=0$ if and only if $a_{-n}$ is sum of derivative.
\label{lmm:74}
\end{lemma}

Then we consider the contribution only from the homogeneous of degree $-n$, for
the residue of the pseudodifferential operator (\ref{eqn:pdo}). 
That is given by the definition of the Wodzicki residue (Def.\ref{def:wzr}).
Furthermore we see the residue is the unique trace.

\begin{theorem}
The noncommutative residue (\ref{eqn:wzr}) is a trace of pseudodifferential operator. If $\dim M>1$, it is the only such trace, up to multiplication by a constant. 
\label{thm:wzr}
\end{theorem}
\proof{Assume that the symbols $a, b$ are supported on a compact subset of a chart domain $U$. The commutator $[A,B]=C$ of pseudodifferential operators corresponds to the composition of the respective symbols $a, b, c$ given by the expansion (\ref{eqn:ppdo})
\begin{align}
c(x,\xi)=\sum_{\alpha\in \NN}\frac{(-i)^{|\alpha|}}{\alpha !}
\left(\partial_\xi^\alpha a\partial_x^\alpha b-\partial_\xi^\alpha b\partial_x^\alpha a\right) .
\end{align}
Each term in this expansion is a finite sum of derivatives, either of the form $\partial p/\partial x^j$ or of the form $\partial q/\partial \xi_i$. In particular, $c_{-n}(x,\xi)=\sum_{j=1}^n \partial p_j/\partial x^j+\partial q_j/\partial \xi_j$, where $p_j$ and $q_j$ are bilinear combinations of the symbols $a$ and $b$ and their derivatives. 
Thus, $\int_{|\xi|=1}\partial p_j/x^j|\sigma_\xi|=\partial P_j/\partial x^j$, where $P_j$ is a smooth function of compact support within $U$; therefore the
integral over $U$ of $\partial P_j/\partial x^j$ vanishes. At the same time, $\int_{|\xi|=1}\partial q_j/\xi^j|\sigma_\xi|=0$
by Lemma\ref{lmm:74}.
Therefore, $\Wres[A,B]=0$.

To verify the uniqueness property, let $T$ be any trace of pseudodifferential operator.
The symbol calculus shows that derivatives are commutators because of
\begin{equation}
[x^j,a]=i\frac{\partial a}{\partial \xi_j},\ \ \  [\xi_j,a]=-i\frac{\partial a}{\partial x^j}.
\label{eqn:poisson}
\end{equation}
Hence $T$ must vanish on derivatives, and thus $T(a)$ depends only on the $(-n)$-homogeneous term $a_{-n}(x,\xi)$.
Consequently, $\Wres$ is the unique trace up to multiplication which is independent of a local coordinate chart.}
\section{Wodzicki residue and heat kernel expansion}
\label{app:heat}

For $A$ a positive elliptic pseudodifferential operator of positive order $m\in\RR$, acting on the space of smooth sections
of an $n$-dimensional vector bundle $E$ over a closed, $n$-dimensional manifold $M$, the zeta function is defined  by eigenvalues $\lambda_j$ as
\begin{align}
\zeta_A(s)=\tr A^{-s}=\sum_j\lambda_j^{-s}, \ \ \ {\cal R}s>\frac{n}{m}\equiv s_0 .
\end{align}
The quotient $s_0=\dim M/{\rm ord}A$ is called the abscissa of convergence of $\zeta_A(s)$, which is proven to have a meromorphic continuation to the whole complex plane $\CC$ (regular at $s_0$), provided that the principal symbol of $A$  admits a spectral cut: $L_\theta=\{\lambda\in \CC, {\rm Arg}\lambda=\theta, \theta_1<\theta<\theta_2\} , {\rm Spec} A\cap L_\theta=\emptyset$ (the Agmon-Nierenberg condition). The precise structure of the analytical continuation of the zeta-function is known in general \cite{HeatKernel}. The only singularities it can have are simple poles at
\begin{align}
s_k=(n-k)/m,\ \ \ k=0,1,2,...,n-1,n+1,...
\end{align}

Now we consider the kernel calculation.
In particular we take $A^{-s}=\dd^{-2s}, s>0$, where one knows the asymptotic behavior of its heat kernel:
\begin{align}
K(x,y;t)&\equiv\frac1{(2\pi)^n}\int e^{i(x-y)\cdot\xi}e^{-tA}d\xi^n,\\
&=\frac1{(4\pi t)^{n/2}}e^{-\sigma(x,y)/2t}\sum_{k=0}b_k(x,y)t^{k/2},\\
K(x,x;t)&=\sum_{k=0}a_k(x,x)t^{(k-n)/2},
\end{align}
for $t\rightarrow 0$.
The explicit form of $a_k(x)$ can be calculated recursively. Here we concern with the spinor representation.
One finds the Seeley-DeWitt coefficients vanish for odd values of $k$ and three of the rest $a_k$'s for $k$ even are\cite{Gilkey};
\begin{align}
a_0(x,x)&= \frac{1}{(4\pi)^{n/2}}\tr (\U)\\
a_2(x,x)&= \frac1{(4\pi)^{n/2}}\tr (-\frac{R}{12})\U \\
a_4(x,x)&= \left(\frac1{(4\pi)^{n/2}}\left(\frac1{288}R^2-\frac1{180}R_{\mu\nu}R^{\mu\nu}-\frac7{1440}R_{\mu\nu\rho\sigma}R^{\mu\nu\rho\sigma}\right.\right.\nonumber\\
&+\left.\left.\frac1{120}\square R
\right)\right)\tr(\U)
-\frac1{12} \tr({\mathbb F}_{\mu\nu}{\mathbb F}^{\mu\nu}) ,
\end{align}
where ${\mathbb F}_{\mu\nu}$ is the bundle curvature or gauge field strength added for further generalization.
(Gilkey 1975\cite{Gilkey} calculated $a_6$; there are 46 possible terms, of which 43 have nonzero coefficients.)
For the calculational aspects and the application to renormalization of quantum field in curved spacetime
(DeWitt, Christensen, Fulling et al. would be referred\cite{etal}) is discussed as the connection between the heat kernel
and the elliptic or hyperbolic Green functions.

Assume $A >c>0$ and from Mellin transform:
\begin{align}
A^{-s}&=\frac1{\Gamma(s)}\int_0^{\infty}e^{-tA}t^s\frac{dt}{t}.
\end{align}
For a pole $s_k=(n-k)/2$, with $s=s_k+\epsilon$,
\begin{align}
kerr_{A^{-s}}(x,x;t)&\equiv\frac1{\Gamma(s)}\int_0^\infty K(x,x;t)t^s \frac{dt}{t}, \\
&=\frac1{\Gamma(s)}\int_0^\infty\sum_{k'} a_{k'}(x,x)t^{(k'-n)/2}t^{s_k+\epsilon} \frac{dt}{t}\ \ ,\\
&=\frac1{\Gamma(s)}\sum a_{k'}(x,x)\int_0^\infty t^{(k'-k)/2+\epsilon}\frac{dt}{t} \label{eqn:43}\\
\res_{s=s_k}kerr_{A^{-s}}(x,x;t)&=\res_{s=s_k}\frac{1}{\Gamma(s_k)}a_k\int_0^\infty t^\epsilon\frac{dt}{t} \ \ .
\end{align}
Since with $\epsilon\rightarrow 0$ the integration is equivalent to
\begin{align}
\int^1_0t^\epsilon\frac{dt}{t}=\frac1{\epsilon},
\end{align}
we determine the relation between the residue and  the coefficient;
\begin{align}
\res_{s=s_k}kerr_{\Delta^-s}(x,x;t)&=\frac{a_k}{\Gamma(s_k)}.
\label{eqn:46}
\end{align}

On the other hand, the kernel of the $\Delta^{-s}$ is derived as follows\cite{Seeley}.
Seeley used the pseudodifferential calculus to construct a good approximation to $(A-\lambda)^{-1}$. Let $B\sim\sum_{k=0}^\infty b_{-d-k}$
\begin{align}
\sigma^{B}\circ \sigma^{A-\lambda}&=1.
\end{align}
From the rule of symbol production,
\begin{align}
b_{-m}(a_m-\lambda)&=1 \\
b_{-m-l}(a_m-\lambda)&+\sum\frac{(-i)^{|\alpha|}}{\alpha !}\partial_\xi^\alpha b_{-m-j}\partial_x^\alpha a_{m-k}=0\ \ \ \ (l>0)
\end{align}
with the sum taken for $j<l, j+k+|\alpha|=l$. They are successively solved for each $b_{-m-l}(x,\xi;\lambda)$,@which is homogeneous of degree $-m-l$.

Now $A^s$ is a pseudodifferential operator with symbol
\begin{align}
\sigma^{A^{-s}}=\sum_{k=0}^\infty \frac1{2\pi i}\oint_{\Gamma}\lambda^s b_{-d-k;\lambda} d\lambda
\end{align}

When ${\cal R}(ms)<-n=-{\rm dim}(M)$,
then $A^s$ has a continuous kernel $k_s(x,y)$ such that $A^s f(x)=\int k_s(x,y) f(y) dy$.
For $x=y$, $k_s(x,x)$ extends to a meromorphic function whose only singularities are simple poles at $s=(j-n)/m,\ \ j=0,1,2...$.
 Each pole is due to a particular term in $\sigma^{A^s}$, and the residue at $s=(j-n)/m$ is
\begin{align}
\res_{s=s_j}k_s(x,x)&=\frac{i}{d(2\pi)^{n+1}}\int_{|\xi|=1}\oint_\Gamma\lambda^s b_{-d-k;\lambda}d\lambda d\xi^{n-1}\\
&=-\frac1{d(2\pi)^n}\int_{|\xi|=1}\sigma_{-n}^{A^{-s}}d\xi^{n-1}.
\label{eqn:52}
\end{align}

From eq.(\ref{eqn:52}) for
\begin{align}
kerr_{A^{-s}}=\frac1{(2\pi)^n}\int e^{i(x-y)\cdot\xi}\sigma^{A^{-s}}d\xi^n,
\end{align}
 the residue at $s=s_k$ is
\begin{align}
\res_{s=s_k} kerr_{A^{-s}}=-\frac1{d(2\pi)^n}\int_{|\xi|=1}\sigma_{-n}^{A^{-s}}d\xi^{n-1}.
\end{align}
Then from eq.(\ref{eqn:46}) we get one of the deepest discoveries of Wodzicki for $\dd^2$
\begin{align}
\wres |\dd|^{k-n}=\frac{(2\pi)^nd}{\Gamma((k-n)/2)}a_k(x,x)|dx^n|.
\end{align}

\section{operator arithmetic}
\label{app:oar}

Operator arithmetic is shortly reviewed.
Firstly we should comment that the definition of derivative about operator has
arbitrarity of ordering by
\begin{align}
\delta F=\overleftarrow{\frac{\partial F}{\partial Q}}\delta Q=\delta Q\overrightarrow{\frac{\partial F}{\partial Q}}
\label{eqn:lrd}
\end{align}
and also their Hermitian ordering is possible.

This multi-definition causes an arbitrariness in the formulation of operator variational principle.
Nevertheless for the traced value $f=\tr F$, a derivative $\partial f/\partial Q$ is uniquely defined because of
\begin{align}
\delta f=\tr \delta F=\tr (\frac{\partial F}{\partial Q}\delta Q)=\tr (\delta Q\frac{\partial F}{\partial Q}).
\end{align}
Then we made up our mind to adopt the derivative from right as our conventional definition:
\begin{align}
\overleftarrow{\frac{\partial F}{\partial Q}}=\frac{\partial F}{\partial Q}.
\end{align}
where apart from $\delta A$, $\frac{\partial F}{\partial A}$ has only a symbolic meaning in this formula; We cannot
determine it as the derived function different from a ordinal function.

As an example of special case, if a operator valued function $F: M_n\mapsto M_n$ is Taylor expanded,
and a function $f$ is given by
\begin{align}
F(A)&=\sum \alpha_iA^i \label{eqn:tay}\\
f(A)&=\tr F(A),
\label{eqn:mfn}
\end{align}
we can easily compute the operator derivative of the function $f$.
\begin{equation}
\delta f =\tr \left(\frac{\partial F}{\partial A}\delta A\right).
\end{equation}
From eq.(\ref{eqn:tay}), it can be reduced to
\begin{align}
\delta f&= \tr \sum_i \sum_{j=0}^{i-1} \alpha_i A^j\delta A A^{i-j-1} \\
&=\tr \sum_i \sum_j \alpha_i A^{i-j-1} A^j\delta A \\
&=\tr \sum_i \alpha_i i A^{i-1}\delta A \\
&=\tr F'(A)\delta A
\end{align}
where $F'(x)$ is a derived function of $F(x)$.

Second example is that
if the function is with matrix valued coefficients, we need more carefulness even in the following simple case,
\begin{align}
F(A)&=BA\\
f(A)&=\tr F(A).
\end{align}

In this case, each ordered derivative is
\begin{align}
\overleftarrow{\frac{\partial F}{\partial A}}&=B\delta A\delta A^{-1} = B,\ \ \ \overrightarrow{\frac{\partial F}{\partial A}}=\delta A^{-1}B\delta A
\end{align}
Even in this non trivial situation, the trace function $f$ can be differentiated as
\begin{align}
\delta f&=\tr\left(\overleftarrow{\frac{\partial F}{\partial A}}\delta A\right)=\tr (B\delta A)\\
&=\tr\left(\delta A\overrightarrow{\frac{\partial F}{\partial A}}\right)=\tr(\delta A\cdot\delta A^{-1}B\delta A)\\
&=\tr\left(\overrightarrow{\frac{\partial F}{\partial A}}\delta A\right)=\tr\left(\delta A\overleftarrow{\frac{\partial F}{\partial A}}\right).
\end{align}

Finally we should note 
\begin{align}
\tr \left(\frac{\partial F}{\partial A}\right) \neq \frac{\partial \left(\tr F\right)}{\partial A}
\end{align}
since the linearity of the trace is maintained as $\delta \tr F=\tr \delta F$.

\end{document}